\documentclass{article}
\pdfoutput=1
\usepackage[T1]{fontenc} 
\usepackage[utf8]{inputenc} 
\usepackage{ismir,amsmath,amssymb,cite,url}
\usepackage{graphicx}
\usepackage{color}
\usepackage{tikz}
\usepackage{subfig}
\usepackage{graphicx}
\usepackage{paralist}
\usepackage{verbatim} 
\usepackage[implicit=false]{hyperref}
\usepackage{multirow}
\usepackage{lineno}

\DeclareMathOperator{\R}{\mathbb{R}}

\title{Music Demixing Challenge 2021}

\author{Yuki Mitsufuji$^1$ \hspace{0.4cm} Giorgio Fabbro$^2$ \hspace{0.4cm} Stefan Uhlich$^2$ \hspace{0.4cm} Fabian-Robert St\"{o}ter$^2$ \\ Alexandre Défossez$^{3}$ \hspace{0.3cm} Minseok Kim$^{4}$ \hspace{0.3cm} Woosung Choi$^{4}$ \hspace{0.3cm} Chin-Yun Yu$^{5}$ \hspace{0.3cm}  Kin-Wai Cheuk$^{6}$\\[0.3cm]
\emph{$^1$ Sony Group Corporation\hspace{0.5cm} $^2$ Sony Europe B.V.\hspace{0.5cm} $^3$ Facebook AI Research Paris} \\
\emph{$^{4}$ Korea University \hspace{0.3cm} $^{5}$ Independent Researcher \hspace{0.3cm} $^{6}$ Singapore University of Technology and Design} \\
{\tt\small Contact: Yuhki.Mitsufuji@sony.com}
}

\begin{document}

\maketitle
\begin{abstract}
Music source separation has been intensively studied in the last decade and tremendous progress with the advent of deep learning could be observed. Evaluation campaigns such as MIREX or SiSEC connected state-of-the-art models and corresponding papers, which can help researchers integrate the best practices into their models. In recent years, the widely used MUSDB18 dataset played an important role in measuring the performance of music source separation. While the dataset made a considerable contribution to the advancement of the field, it is also subject to several biases resulting from a focus on Western pop music and a limited number of mixing engineers being involved. To address these issues, we designed the Music Demixing~(MDX) Challenge on a crowd-based machine learning competition platform where the task is to separate stereo songs into four instrument stems~(Vocals, Drums, Bass, Other). The main differences compared with the past challenges are 1) the competition is designed to more easily allow machine learning practitioners from other disciplines to participate, 2) evaluation is done on a hidden test set created by music professionals dedicated exclusively to the challenge to assure the transparency of the challenge, i.e., the test set is not accessible from anyone except the challenge organizers, and 3) the dataset provides a wider range of music genres and involved a greater number of mixing engineers. In this paper, we provide the details of the datasets, baselines, evaluation metrics, evaluation results, and technical challenges for future competitions.
\end{abstract}
\section{Introduction}\label{sec:introduction}
Audio source separation has been studied extensively for decades as it brings benefits in our daily life, driven by many practical applications, e.g., hearing aids, denoising in video conferences, etc. Additionally, music source separation (MSS) attracts professional creators because it enables the remixing or reviving of songs to a level, never achieved with conventional approaches such as equalizers. Further, suppressing vocals in songs can improve the experience of a karaoke application, where people can enjoy singing together on top of the original song (where the vocals were suppressed), instead of relying on content developed specifically for karaoke applications. Despite the potential benefits, the research community struggled to achieve a separation quality required by commercial applications. These demanding requirements were also aggravated by the under-determined settings the problem was formulated in since the number of provided channels in the audio recording is less than the number of sound objects that need to be separated.

In the last decade, the separation quality of MSS has mainly been improved owing to the advent of deep learning. A significant improvement in an MSS task was observed at the Signal Separation Evaluation Campaign~(SiSEC) 2015~\cite{SiSEC2015}, where a simple feed-forward network~\cite{7178348} able to perform four-instruments separation achieved the best signal-to-distortion ratio~(SDR) scores, surpassing all other methods that did not use deep learning. The use of deep learning in MSS was accelerated ever since and led to improved SDR results year after year in the successive SiSEC editions, held in 2016~\cite{SiSEC2016} and 2018~\cite{SiSEC2018}. 
An important component of this success story was the release of publicly available datasets such as~\cite{MUSDB18} which, compared to previous datasets such as~\cite{bittner2014medleydb}, was created specifically for MSS tasks.
MUSDB18 consists of 150 music tracks in four stems and is up until now widely used due to a lack of alternatives\footnote{An overview of available dataset for music source separation is given in https://source-separation.github.io/tutorial/data/datasets.html}.
The dataset also has a number of limitations such as its limited number of genres (mostly pop/rock) and its biases concerning mixing characteristics (most stems were produced by the same engineers).
Since the last evaluation campaign took place, many new papers were published claiming state-of-the-art, based on MUSDB18 test data, however, it is unclear if generalization performance did improve at the same pace or if some models overfit on MUSDB18.
To keep scientific MSS research relevant and sustainable, we want to address some of the limitations of current evaluation frameworks by using: 

\begin{itemize}
\item a fully automatic evaluation system enabling straightforward participation for machine learning practitioners from other disciplines.
\item a new professionally produced dataset containing unseen data dedicated exclusively to the challenge to ensure transparency in the competition (i.e., the test set is not accessible from anyone except the challenge organizers).
\end{itemize}

With these contributions, we designed a new competition called Music Demixing~(MDX) Challenge\footnote{https://www.aicrowd.com/challenges/music-demixing-challenge-ismir-2021}, where a call for participants was conducted on a crowd-based machine learning competition platform. A hidden dataset crafted exclusively for this challenge was employed in a system that automatically evaluated all MSS systems submitted to the competition. The MDX Challenge is regarded as a follow-up event of the professionally-produced music~(MUS) task of the past SiSEC editions; to continue the tradition of the past MUS, participants were asked to separate stereo songs into stems of four instruments~(Vocals, Drums, Bass, Other). Two leaderboards are used to rank the submissions: A) methods trained on MUSDB18(-HQ) and B) methods trained with extra data.
Leaderboard A gives the possibility to any participant, independently on the data they possess, to train a MSS system (since MUSDB18 is open) and includes systems that can, in a later stage, be compared with the existing literature, as they share the same training data commonly used in research; leaderboard B permits models to be used to their full potential and therefore shows the highest achievable scores as of today.

In the following, the paper provides the details about the test dataset in Sec.~\ref{Dataset}, the leaderboards in Sec.~\ref{Leaderboard}, the evaluation metrics in Sec.~\ref{Eval_Metric}, the baselines in Sec.~\ref{Baselines}, the evaluation results in Sec.~\ref{Eval_Result}, and the technical challenges for future competitions in Sec.~\ref{Future}.

\section{MDXDB21}
\label{Dataset}
For the specific purpose of this challenge, we introduced a new test set, called MDXDB21.
This test set is made of 30 songs, created by Sony Music Entertainment~(Japan) Inc.~(SMEJ) with the specific intent to use it for the evaluation of the MDX Challenge.
The dataset was hidden from the participants, only the organizers of the challenge could access it.
This allowed a fair comparison of the submissions. Here we provide details on the creation of the dataset:
\begin{itemize}
\item More than 20 songwriters were involved in the making of the 30 songs in the dataset, so that there is no overlap with existing songs in terms of composition and lyrics;
\item The copyright of all 30 songs is managed by SMEJ so that MDXDB21 can be integrated easily with other datasets in the future, without any issue arising from copyright management;
\item More than 10 mixing engineers were involved in the dataset creation with the aim of diversifying the mixing styles of the included songs;
\item The loudness and tone across different songs were not normalized to any reference level, since these songs are not meant to be distributed on commercial platforms;
\item To follow the tradition of past competitions like SiSEC, the mixture signal (i.e., the input to the models at evaluation time) is obtained as the simple summation of the individual target stems.
\end{itemize}

Table~\ref{tab:MDXDB21} shows a list of the songs included in MDXDB21.
To give more diversity in genre and language, the dataset features also non-western and non-English songs.
The table also provides a list of the instruments present in each song; this can help researchers understand under which conditions their models fail to perform the separation. More information about the loudness as well as stereo information for each song and its stems are given in the Appendix.

\begin{table*}
 \begin{center}
 \resizebox{\linewidth}{!}{
 \begin{tabular}{|l|l|l|l|l|}
  \hline
Song ID & Genre & Language & Title & Instruments in \it{Other}\\
  \hline
    \hline
  SS\_001  & Jpop & Japanese & Give Me One More Day & ch, db, egtr, gtr, hn, perc, pf, sfx, va, vc, vn\\
  \hline
  SS\_002  & Acid Jazz & English & Do It Right & egtr, epf, fl, perc, pf, sfx, str\\
  \hline
  SS\_003  & Ballad & English & I Reach for You & ch, pf, pk, va, vc, vn\\
  \hline
  SS\_004  & Funk & English & Diamond In a Rough & asax, gl, gtr, hpd, key, perc, syn, tbn, tpt, \\
  & & & & tsax, va, vc, vn\\  
  \hline
  SS\_005  & Ballad~(Duet) & English & You and I & cl, db, fl, gl, hn, hp, ob, pf, va, vc, vn\\
  \hline
  SS\_006  & Jazz & English & City Swing & asax, gtr, key, pf, tbn, tpt, tsax\\
  \hline
  SS\_007  & Bigband & English & Cut Loose, Then Ya Vamoose & asax, clap, gtr, key, tpt, tbn, tsax\\
  \hline
  SS\_008  & Rock & English & Take You Out & gtr, perc, pf\\
  \hline
  SS\_009  & Country & English & For Your Smile & egtr, gtr, mand, org, pad, perc, pf, sfx\\
  \hline
  SS\_010  & Hiphop & English & Monsta & bell, sfx, syn\\
  \hline
  SS\_011  & EDM & English & Tormented Soul & bell, egtr, hpd, perc, pf, sfx\\
  \hline
  SS\_012  & Metal & English & Time to Let Go & gtr, key\\
  \hline
  SS\_013  & Jazz & English & This Is Day & bell, gtr, pad, pf, sfx, syn\\
  \hline
  SS\_014  & AOR & English & Moments & gtr, mar, perc, syn, xyl\\
  \hline
  SS\_015  & Indian Pop & Hindi & Dil Ki Baatein & syn, egtr\\
   &  & & (Conversations of the Heart) & \\
  \hline
  SS\_016  & Healing & Japanese & Waraeya Utaeya & epf, pad, perc, pf, str, shakuhachi, sfx, syn\\
  \hline
  SS\_017  & Reggae & English & Jump up & ch, gtr, org, perc, pf\\
  \hline
  SS\_018  & Pop & English & I Wanna See & gl, gtr, pad, pf, syn, vib, va, vc, vn\\
  \hline
  SS\_019  & Pop & English & Tonight & bell, egtr, epf, pad, sfx, str, syn\\
  \hline
  SS\_020  & Rock & English & Six Feet Between & org, pad, pf, syn\\
  \hline
  SS\_021  & R\&B & Portuguese & Minha Sina & epf, pad, pf\\
  \hline
  SS\_022  & Jpop & Japanese & Spring & bell, egtr, gl, koto, perc, pf, sfx, \\
  & & & (Three hundred years later) & shakuhachi, syn, va, vc, vn\\  
  \hline
  SS\_023  & Jpop & Japanese & Scissorhands & egtr, gtr, perc\\
  \hline
  SS\_024  & Digital Rock & English & I Can See You Calling Me & pad, sfx, syn\\
  \hline
  SS\_025  & Kpop & Korean & We Can Fly Away & pf, sfx, syn\\
  \hline
  SS\_026  & Electro & English & A.I. Robot & epf, gtr, pad, perc, pf, sfx, str, syn\\
  \hline
  SS\_027  & Trance & English & By My Side & gtr, pad, sfx, syn\\
  \hline
  SS\_028  & House & English & Break Free~(Dance Together) & perc, pf, syn\\
  \hline
  SS\_029  & Pop & English & By Your Side & gtr, pad, pf, syn\\
  \hline
  SS\_030  & Jpop & Japanese & Kikai Jikake no Koenaki Katsubou & pf, sfx, str, syn\\
  \hline
 \end{tabular}}
\end{center}
 \caption[List of songs in MDXDB21. Abbreviations of the instrument names comply with the International Music Score Library Project (IMSLP)]{List of songs in MDXDB21. Abbreviations of the instrument names comply with the International Music Score Library Project (IMSLP)\protect\footnotemark.}
 \label{tab:MDXDB21}
\end{table*}

\footnotetext{\textcolor{black}{https://imslp.org/wiki/IMSLP:Abbreviations\_for\_Instruments}}

\section{Leaderboards and Challenge Rounds}
\label{Leaderboard}
For a fair comparison between systems trained with different data, we designed two different leaderboards for the MDX Challenge:

\begin{itemize}
    \item \textbf{Leaderboard A}
    accepted MSS systems that are trained exclusively on MUSDB18-HQ~\cite{MUSDB18HQ}\footnote{Participants were encouraged to use MUSDB18-HQ as opposed to MUSDB18, as the former is not limited to 16kHz, a limit imposed by the audio codec of MUSDB18. Nevertheless, systems trained on MUSDB18 were also eligible for leaderboard A as MUSDB18 can be seen as a ``derived'' version of MUSDB18-HQ.}.
    Our main purpose was to give everyone the opportunity to start training a MSS model and take part in the competition, independently on the data they have. On top of that, since MUSDB18-HQ is the standard training dataset for MSS in literature, models trained with it can also be compared with the current state-of-the-art in publications, by evaluating their performance on the test set of MUSDB18-HQ and using the metrics included in the BSS Eval v4 package, as done for example by \cite{defossez2021hybrid} and \cite{kim2021kuielab}.

    \item \textbf{Leaderboard B}
did not pose any constraints on the used training dataset. This allowed participants to train bigger models, exploiting the power of all the data at their disposal.
\end{itemize}

To avoid some participants overfitting to the MDXDB21 dataset, we split the dataset into three equal-sized parts and designed two challenge rounds: in the first round, participants could access the scores of their models computed only on the first portion of MDXDB21. In the second round, the second portion of MDXDB21 was added to the evaluation and participants could see how well their models generalized on new data. After the challenge ended, the overall score was computed on all songs\footnote{\label{refdemo}We excluded three songs, which we used for demo purposes, e.g., to provide feedback to the participants about their submissions on AIcrowd. These three songs are SS\_008, SS\_015 and SS\_018.}. These overall scores were also used for the final ranking of the submissions.

\section{Evaluation Metric}
\label{Eval_Metric}
In the following section, we will introduce the metric that was used for the MDX Challenge as the ranking criterion and compare it to other metrics that have been used in past competitions.

\subsection{Signal-to-Distortion Ratio (SDR)}
As an evaluation metric, we chose the \textcolor{black}{multichannel} signal-to-distortion ratio~(SDR) \cite{vincent2007first}, also called signal-to-noise ratio~(SNR), which is defined as
\begin{equation}
SDR = 10 \log_{10} \frac{\sum_{n} \lVert\mathbf{s}(n)\rVert^2 + \epsilon}{\sum_{n}\lVert\mathbf{s}(n)-\hat{\mathbf{s}}(n)\rVert^2 + \epsilon},
\label{eq:sdr}
\end{equation}
where $\mathbf{s}(n) \in \R^2$ denotes the waveform of the ground truth and $\hat{\mathbf{s}}(n)$ the waveform of the estimate for one of the sources with $n$ being the (discrete) time index.
We use a small constant $\epsilon = 10^{-7}$ in \eqref{eq:sdr} to avoid divisions by zero.
The higher the SDR score is, the better the output of the system is.

For each song, this allows computing the average SDR, $SDR_{\rm{Song}}$, given by
\begin{multline}
SDR_{\rm{Song}} = \frac{1}{4}\biggl(SDR_{\rm{Bass}} \\ + SDR_{\rm{Drums}} + SDR_{\rm{Other}} + SDR_{\rm{Vocals}}\biggr).
\end{multline}
Finally, the systems are ranked by averaging $SDR_{\rm{Song}}$ over all songs in the hidden test set.

As \eqref{eq:sdr} considers the full audio waveform at once, we will denote it as a ``global'' metric. In contrast, we will denote a metric as ``framewise'' if the waveform is split into shorter frames before analyzing it. Using the global metric \eqref{eq:sdr} has two advantages. First, it is not expensive to compute as opposed to more complex measures like BSS Eval v4 which also outputs the image-to-spatial distortion ratio (ISR),  signal-to-interference ratio (SIR), and signal-to-artifacts ratio (SAR). Second, there is also no problem with frames where at least one source or estimate is all-zero. Such frames are discarded in the computation of BSS Eval v4 as otherwise, e.g., SIR can not be computed. This, however, yields the unwanted side-effect that the SDR values of the different sources of BSS Eval v4 are not independent of each other which is not desired\footnote{See for example\\ \texttt{https://github.com/sigsep/bsseval/issues/4}.}. The global SDR \eqref{eq:sdr} does not suffer from this cross-dependency between source estimates.

\subsection{Comparison with Other Metrics}
Before deciding to choose the global SDR~\eqref{eq:sdr}, we did a comparison with other metrics for audio source separation for the best system from SiSEC 2018 (``TAU1''). Other common metrics are
\begin{itemize}
    \item[(a)] Global/framewise SDR,
    \item[(b)] Global/framewise SI-SDR~\cite{le2019sdr},
    \item[(c)] Global/framewise mean absolute error (MAE),
    \item[(d)] Global/framewise mean squared error (MSE),
    \item[(e)] SDR of multi-channel BSS Eval v3\\(evaluation metric of SiSEC 2015)~\cite{vincent2007first,vincent2012signal,SiSEC2015},\footnote{This metric is equivalent to \eqref{eq:sdr} except for the small constant $\epsilon$.}
    \item[(f)] Mean/median of framewise multi-channel BSS Eval~v3\\(evaluation metric of SiSEC 2016)~\cite{SiSEC2016},
    \item[(g)] Mean/median of framewise multi-channel BSS Eval v4\\(evaluation metric of SiSEC 2018, available as \texttt{museval} Python package)~\cite{SiSEC2018}.
\end{itemize}
``Global'' refers to computing the metric on the full song whereas ``framewise'' denotes a computation of the metric on shorter frames which are then averaged to obtain a value for the song.
For the framewise metrics, we used in our experiment a frame size as well as a hop size of one second, which is the default for \texttt{museval}~\cite{SiSEC2018}, except for the framewise SDR of SiSEC 2016 where we used a frame size of 30 seconds and a hop size of 15 seconds as in \cite{SiSEC2016}.

\begin{figure*}
    \centering
    \resizebox{0.7\linewidth}{!}{    \includegraphics{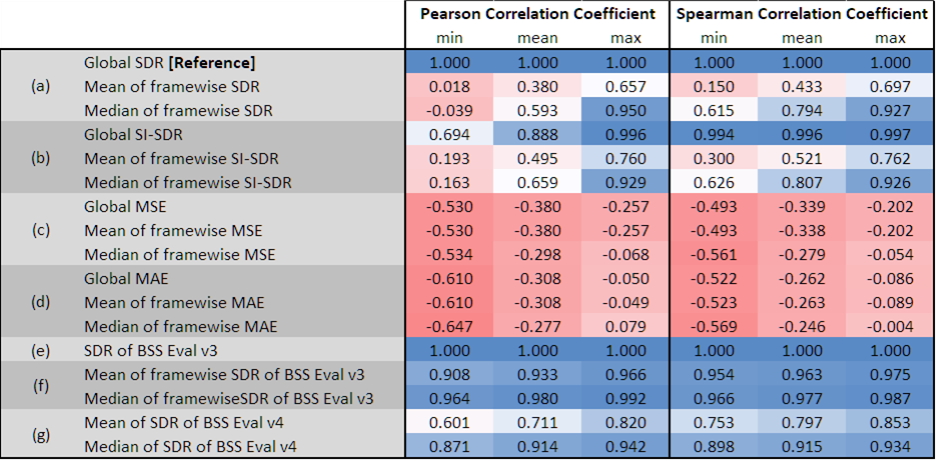}}
    \caption{Comparison of MSS metrics using Pearson and Spearman correlation. Metrics are compared to ``median of framewise SDR'', i.e., to the metric that was used for SiSEC 2018.}
    \label{fig:correlation_analysis}
\end{figure*}

Fig.~\ref{fig:correlation_analysis} shows the correlations of the global SDR \eqref{eq:sdr} (used as reference) with the different metrics on MUSDB18 Test for ``TAU1'', the best system from SiSEC 2018 \cite{SiSEC2018}.
For each metric, we compute the correlation coefficient for each source to the reference metric and show the minimum, average, and maximum correlation over all four sources. \textcolor{black}{Please note that some metrics are similarity metrics (``higher is better'') whereas others measure the distance (``smaller is better''). As we use the global SDR as reference, the correlation coefficient becomes negative if the correlation with a distance metric is computed.}
We can see that there is a strong correlation between the used global SDR \eqref{eq:sdr} and the median-averaged SDR from BSS Eval v3 and v4 as the Pearson and Spearman correlations are on average larger than $0.9$. This analysis confirms that the global SDR \eqref{eq:sdr} is a good choice as it has a high correlation
to the evaluation metrics of SiSEC 2016, i.e., metric (f), and SiSEC 2018, i.e., metric (g), while being at the same time simple to compute and yielding per-source values which are independent of the estimates for other sources. In the following, we will refer to the global SDR \eqref{eq:sdr} as ``SDR''.

\section{Baseline Systems}
\label{Baselines}
The MDX Challenge featured two baselines: Open-Unmix (UMX) and CrossNet-UMX (X-UMX).
A description of UMX can be found in~\cite{Stoter2019}, where the network is based on a BiLSTM architecture that was studied in~\cite{Uhlich17}.
X-UMX~\cite{sawata2021all} is an enhanced version of UMX.

\begin{figure*}[th]
    \begin{minipage}{.49\linewidth}
        \centering
        \includegraphics[scale=0.45]{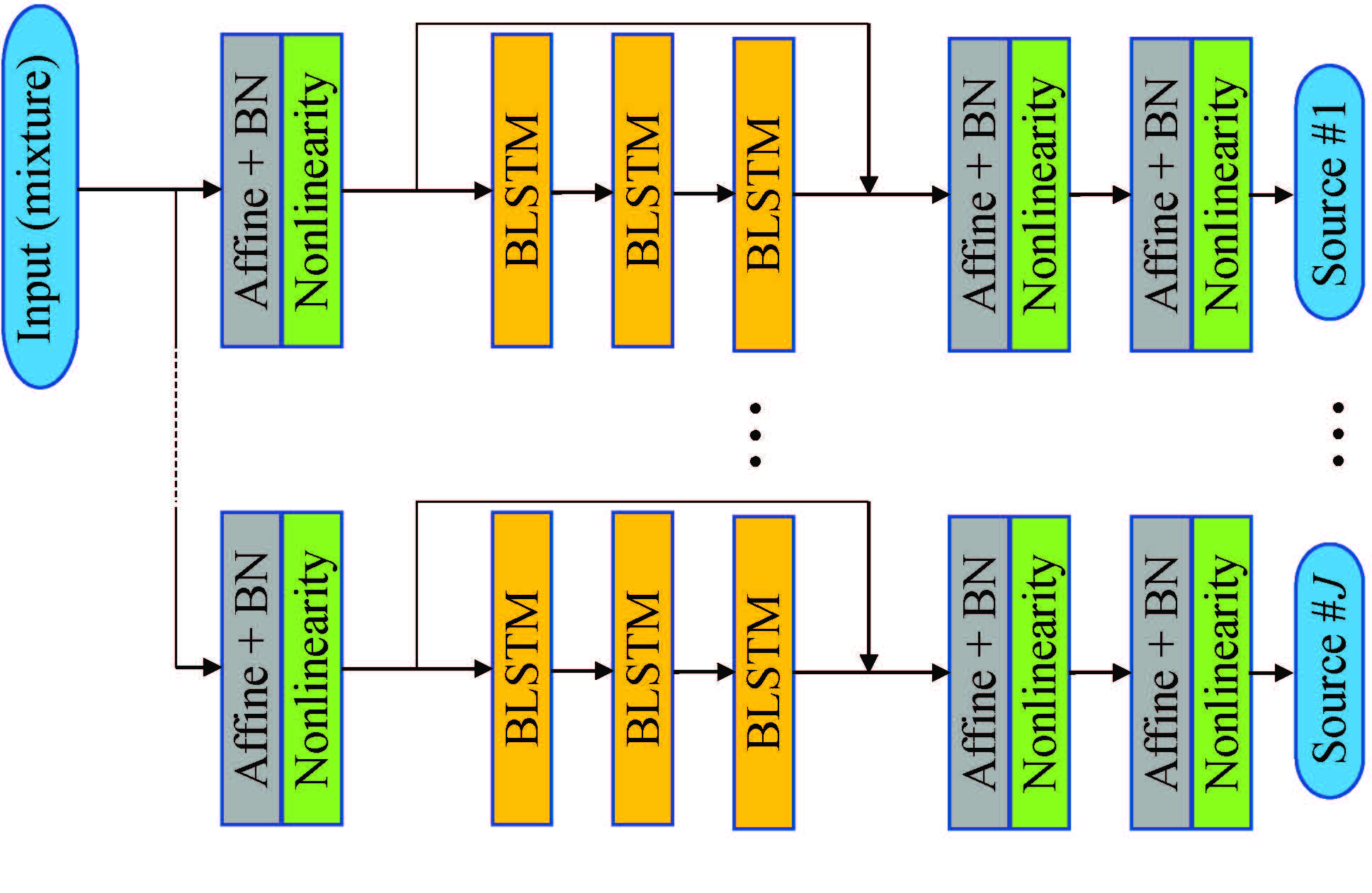}

        \small (a) UMX
    \end{minipage}
    \begin{minipage}{.49\linewidth}
        \centering
        \includegraphics[scale=0.45]{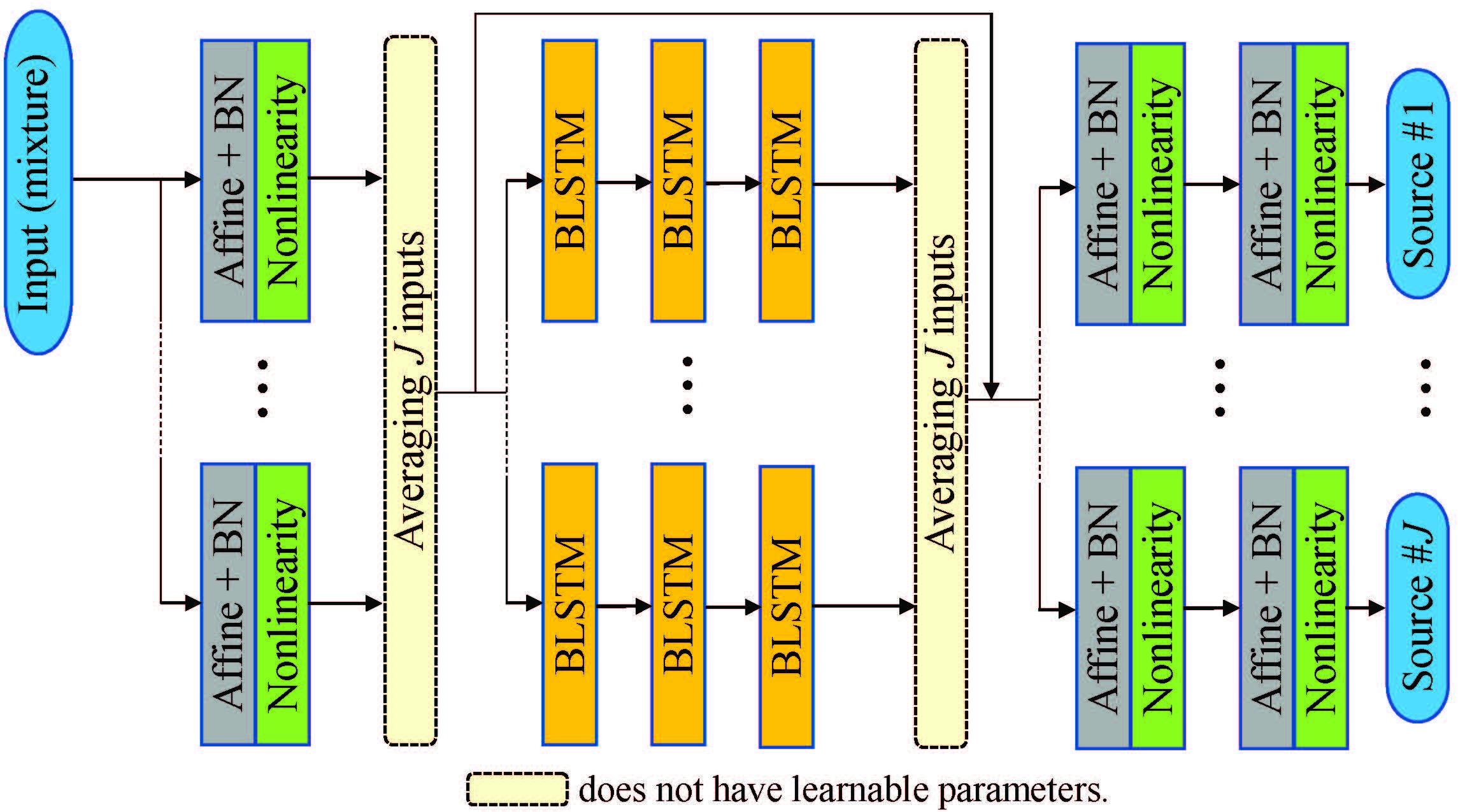}
        
        \small (b) X-UMX
    \end{minipage}
    
    \caption{Comparison of network architectures used in our experiments.}
  \label{fig:comp_archi}
\end{figure*}

Fig.~\ref{fig:comp_archi} shows the architectures of the two models. The main difference between them is that UMX can be trained independently for any instrument while X-UMX requires all the networks together during training to allow the exchange of gradients between them, at the cost of more memory. During inference, there is almost no difference between UMX and X-UMX regarding the model size or the computation time as the averaging operations in X-UMX do not introduce additional learnable parameters.

\section{MDX Challenge 2021 Results and Key Takeaways}
\label{Eval_Result}
In this section, we will first give results for various systems known from the literature on MDXDB21 before summarizing the outcome of the MDX Challenge 2021.

\subsection{Preliminary Experiments on MDXDB21}
The two baselines described in Sec.~\ref{Baselines}, as well as state-of-the-art MSS methods, were evaluated on MDXDB21. Table~\ref{tab:result_LBA} shows SDR results averaged over all 27 songs on leaderboard A, where all the listed models were trained only either on MUSDB18 or on MUSDB-HQ. For leaderboard B, UMX and X-UMX were trained on both, training and test set of MUSDB18-HQ, using the same 14 songs for validation as if only the train part of MUSDB18 would have been used. Table~\ref{tab:result_LBB} shows the SDR results on leaderboard B. Since the extra dataset used in each model is different, we cannot directly compare their scores, but we nonetheless listed the results to see how well the SOTA models can perform on new data created by music professionals. It can already be seen that the difference in SDR between these models is smaller than what is reported on the well-established MUSDB18 test set. This indicates that generalization is indeed an issue for many of these systems and it will be interesting to see if other contributions can outperform the SOTA systems.

\begin{figure*}
    \centering
    \resizebox{\linewidth}{!}{\includegraphics{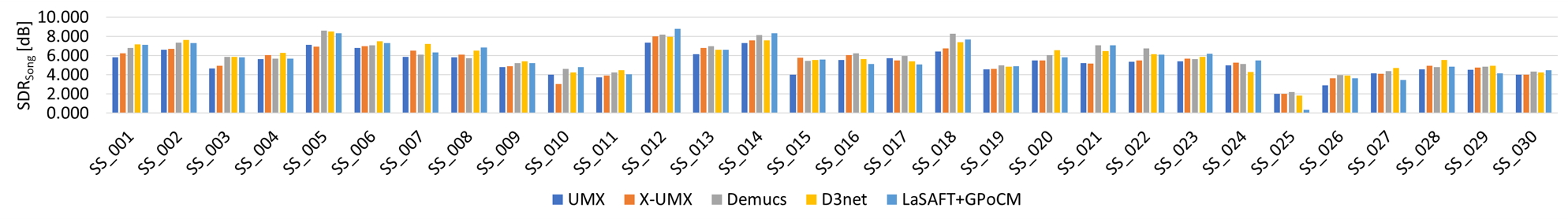}}
    \small (a) Leaderboard A

    \resizebox{\linewidth}{!}{\includegraphics{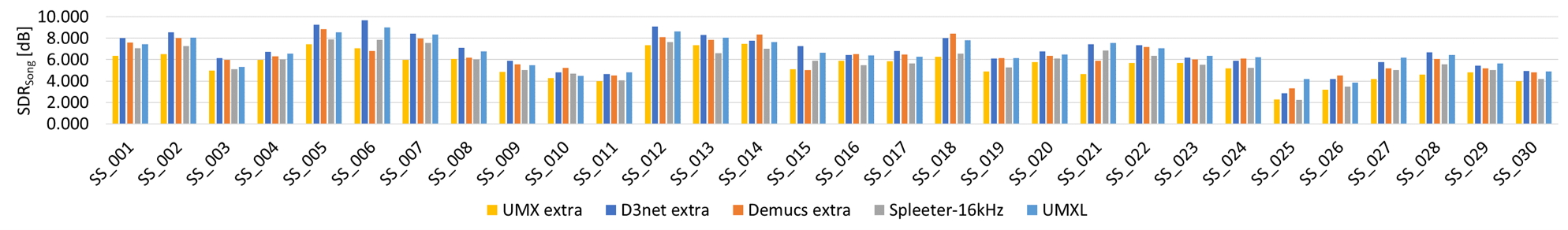}}
    \small (b) Leaderboard B

    \caption{Individual $SDR_{\rm{Song}}$ for methods known from literature for leaderboard A and B.}
    \label{fig:leaderboard}
\end{figure*}

Fig.~\ref{fig:leaderboard}(a) and \ref{fig:leaderboard}(b) show $SDR_{\rm{song}}$ for all 30 songs for the baselines as well as the currently best methods known from literature. There is one exception in the computation of $SDR_{\rm{song}}$ for SS\_015: For this song, it was computed by excluding $\it{bass}$ and averaging over three instruments only as this song has an all-silent $\it{bass}$ track and $SDR_{\rm{Bass}}$ would aggravate the average SDR over four instruments. \textcolor{black}{For this reason this song was made part of the set of three songs (SS\_008, SS\_015 and SS\_018) that were left out of the evaluation so that they could be provided to the participants as demo (see note \ref{refdemo}).}

The results for SS\_025--026 are considerably worse than for the other songs; we assume that this is because these songs contain more electronic sounds than the others. 

\begin{table*}[t]
 \centering
 \resizebox{\linewidth}{!}{
 \begin{tabular}{|l|l|l|l|l|l|l|}
  \hline
  System & Training Set & $SDR_{\rm{Song}}$ & $SDR_{\rm{Bass}}$ & $SDR_{\rm{Drums}}$ & $SDR_{\rm{Other}}$ & $SDR_{\rm{Vocals}}$ \\
  \hline
  \hline
  Ideal SWF & \textit{Oracle} & 9.63 &   9.01 & 9.54 & 8.63 & 11.33 \\
  Ideal MWF & \textit{Oracle} & 9.78 & 9.39 & 9.59 & 8.84 & 11.30 \\
  \hline
  UMX~\cite{Stoter2019} & MUSDB18-HQ & 5.18 & 5.40 & 5.71 & 3.56 & 6.07 \\
  X-UMX~\cite{sawata2021all} & MUSDB18-HQ & 5.37 & 5.62	& 5.81 & 3.72 & 6.34 \\
  \hline
  LaSAFT+GPoCM~\cite{choi2021lasaft} & MUSDB18 & 5.64 & 5.78 & 5.23 & 4.37 & 7.16 \\
  D3Net~\cite{takahashi2021d3net} & MUSDB18 & 5.80 & 5.74 & 6.18 & 4.30 & 6.97\\
  Demucs~\cite{defossez2021music} & MUSDB18 & 5.81 & 6.48 & 6.44 & 3.96 & 6.37\\
  \hline
 \end{tabular}}
 \caption{SDR results on MDXDB21 for oracle baselines as well as systems known from literature which are eligible for leaderboard A, i.e., systems trained only on MUSDB18/MUSDB18-HQ.}
 \label{tab:result_LBA}
\end{table*}

\begin{table*}[t]
 \centering
 \resizebox{\linewidth}{!}{
 \begin{tabular}{|l|l|l|l|l|l|l|l|}
  \hline
  System & Training Set & $SDR_{\rm{Song}}$ & $SDR_{\rm{Bass}}$ & $SDR_{\rm{Drums}}$ & $SDR_{\rm{Other}}$ & $SDR_{\rm{Vocals}}$ \\
  \hline
  \hline
  UMX extra~\cite{Stoter2019} & MUSDB18-HQ + 50 songs & 5.42 & 5.60 & 5.94 & 3.75 & 6.36 \\
  X-UMX extra~\cite{sawata2021all} & MUSDB18-HQ + 50 songs & 5.59 & 5.79 & 6.26 & 3.90 & 6.41 \\
  \hline  
  Demucs extra~\cite{defossez2021music} & MUSDB18 + 150 songs & 6.33 & 6.96 & 6.92 & 4.51 & 6.94 \\
  D3Net extra~\cite{takahashi2021d3net} & MUSDB18 + 1.5k songs & 6.67 & 6.58 & 6.82 & 5.24 & 8.06\\
  Spleeter (11kHz)~\cite{Hennequin2020} & 25k songs & 5.72 & 5.77 & 5.85 & 4.32 & 6.94 \\
  Spleeter (16kHz)~\cite{Hennequin2020} & 25k songs & 5.77 & 5.77 & 5.97 & 4.33 & 6.99 \\
  UMXL~\cite{Stoter2019} & 200h MP3 songs & 6.52 &  6.62 & 6.84 & 4.89 & 7.73 \\
  \hline
 \end{tabular}}
 \caption{SDR results on MDXDB21 for methods known from literature which are eligible for leaderboard~B, i.e., systems allowed to train with extra data.}
 \label{tab:result_LBB}
\end{table*}

\subsection{MDX Challenge 2021 Results}

\begin{figure*}
    \centering
    \resizebox{\linewidth}{!}{\includegraphics{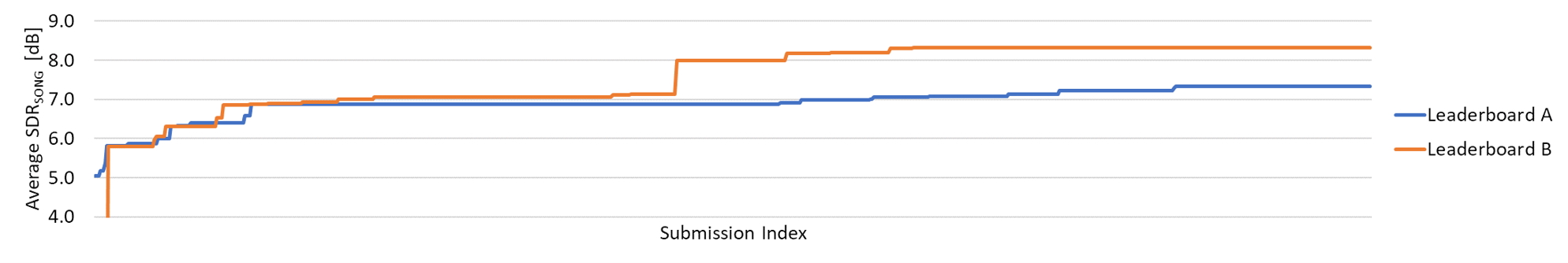}}
    \caption{Evolution of best $SDR_\text{Song}$ over time (for leaderboard A and B).}
    \label{fig:sdr_over_time}
\end{figure*}

The challenge was well received by the community and, in total, we received 1541 submissions from 61 teams around the world.
Table~\ref{tab:MDX_final_results} shows the final leaderboards of the MDX Challenge. \textcolor{black}{It gives the mean SDR from \eqref{eq:sdr} for all 27 songs as well as the standard-deviation for the mean SDR over the three splits (each with 9 songs) that were used in the different challenge rounds as described in Sec~\ref{Leaderboard}.} Comparing these numbers with the results in Table~\ref{tab:result_LBA} and \ref{tab:result_LBB}, we can observe a considerable SDR improvement of approximately 1.5dB throughout the contest. The evolution of the best SDR over time is shown in Fig.~\ref{fig:sdr_over_time}. \textcolor{black}{As several baselines for leaderboard A were provided at the start of the challenge, progress could be first observed for leaderboard A and the submissions for leaderboard B were not significantly better. With the start of round 2, the performance gap between leaderboard A and leaderboard B increased as participants managed to train bigger models with additional data.}

\begin{table*}
\centering
\resizebox{\linewidth}{!}{
\begin{tabular}{|l|l|l|l|l|l|l|l|}
    \hline
     Rank & Team/User name & Model name & $SDR_{\rm{Song}}$ & $SDR_{\rm{Bass}}$ & $SDR_{\rm{Drums}}$ & $SDR_{\rm{Other}}$ & $SDR_{\rm{Vocals}}$ \\
  \hline
  \hline
  1. & defossez               & Hybrid Demucs   & 7.328$\pm$0.46 & 8.115$\pm$0.54 & 8.037$\pm$0.36 & 5.193$\pm$1.02 & 7.968$\pm$1.03 \\
  2. & kuielab                & KUIELab-MDX-Net & 7.236$\pm$0.59 & 7.232$\pm$0.42 & 7.173$\pm$0.46 & 5.636$\pm$0.89 & 8.901$\pm$1.41 \\
  3. & Music\_AI              & -               & 6.882$\pm$0.23 & 7.273$\pm$0.37 & 7.371$\pm$0.62 & 5.091$\pm$0.73 & 7.792$\pm$0.86 \\
  4. & Kazane\_Ryo\_no\_Danna & Danna-Sep       & 6.649$\pm$0.41 & 6.993$\pm$0.56 & 7.018$\pm$0.58 & 4.901$\pm$0.91 & 7.686$\pm$1.15 \\
  5. & ByteMSS                & -               & 6.514$\pm$0.48 & 6.602$\pm$0.42 & 6.545$\pm$0.48 & 4.830$\pm$0.86 & 8.079$\pm$1.24 \\
  \hline
\end{tabular}}

\vspace{0.2cm}

\scriptsize (a) Leaderboard A

\vspace{0.3cm}
\resizebox{\linewidth}{!}{
\begin{tabular}{|l|l|l|l|l|l|l|l|}
    \hline
     Rank & Team/User name & Model name & $SDR_{\rm{Song}}$ & $SDR_{\rm{Bass}}$ & $SDR_{\rm{Drums}}$ & $SDR_{\rm{Other}}$ & $SDR_{\rm{Vocals}}$ \\
  \hline
  \hline
  1. & Audioshake & - & 8.326$\pm$0.44 & 8.342$\pm$0.45& 8.664$\pm$0.37 & 6.505$\pm$0.85 &	9.793$\pm$1.13 \\
  2. & defossez   & Hybrid Demucs & 8.110$\pm$0.27 &	8.856$\pm$0.65 &	8.850$\pm$0.22 &	5.978$\pm$0.72 &	8.756$\pm$0.95\\
  3. & kuielab    & KUIELab-MDX-Net & 7.370$\pm$0.47 &	7.495$\pm$0.18 &	7.554$\pm$0.57 &	5.533$\pm$0.82 &	8.896$\pm$1.30\\
  4. & Music\_AI  & - & 6.882$\pm$0.23 & 7.273$\pm$0.37 & 7.371$\pm$0.62 & 5.091$\pm$0.73 &	7.792$\pm$0.86\\
  5. & PyMDX      & - & 6.680$\pm$0.32 & 6.619$\pm$0.56 & 6.838$\pm$0.46 & 4.891$\pm$0.75 &	8.374$\pm$1.07\\
  \hline
\end{tabular}}

\vspace{0.2cm}

\scriptsize (b) Leaderboard B
\caption{Final leaderboards with top-5 submissions evaluated on MDXDB21.}
\label{tab:MDX_final_results}
\end{table*}

This improvement was not only achieved by blending several existing models but also by architectural changes. The following sections give more details for the winning models and they are written by each team, respectively.

\subsection{Hybrid Demucs (defossez)}
\label{ssec:defossez}

\begin{figure*}
    \centering
    \includegraphics[width=0.85\textwidth]{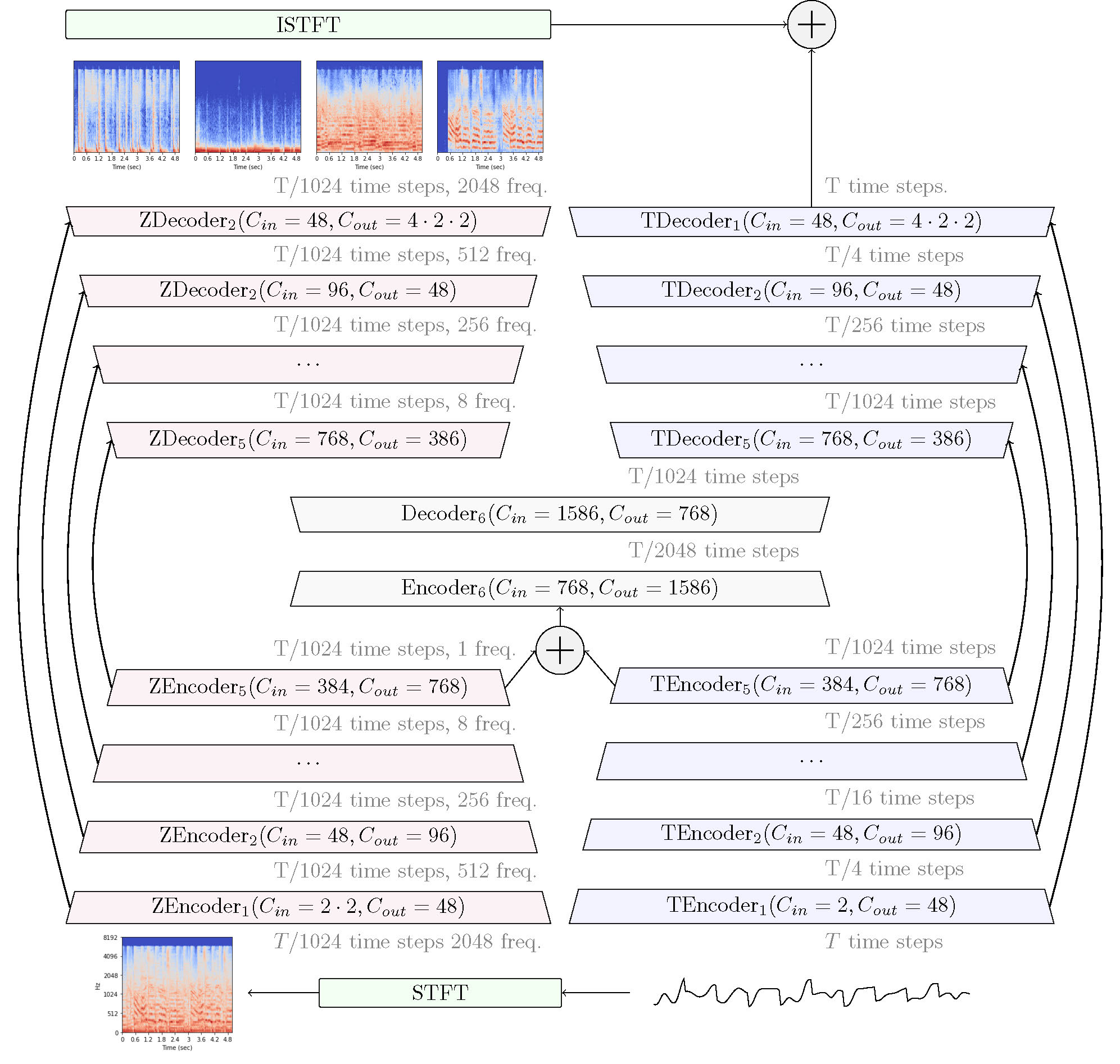}
    \caption{Hybrid Demucs architecture. The input waveform is processed
both through a temporal encoder and through the STFT followed by
a spectral encoder. After layer 5, the two representations have the same shape and are summed before going through shared layers.
The decoder is built symmetrically. The output spectrogram go through the ISTFT
and is summed with the waveform outputs, giving the final model output.
The $\mathrm{Z}$ prefix is used for spectral layers, and $\mathrm{T}$ prefix for the
temporal ones.}
    \label{fig:hybrid_demucs}
\end{figure*}

Hybrid Demucs extends the original Demucs architecture~\cite{defossez2021music} with multi-domain analysis and prediction
capabilities. The original model consisted of an encoder/decoder in the time domain, with U-Net skip connection~\cite{unet}.
In the hybrid version, a spectrogram branch is added, which is fed with the input spectrogram, either represented by its amplitude,
or real and imaginary part, a.k.a Complex-As-Channels (CAC)~\cite{choi_2020}. Unlike the temporal branch, the spectrogram branch applies
convolutions along the frequency axis, reducing the number of frequency bins by a factor of 4 with every encoder layer.
Starting from 2048 frequency bins, obtained with a 4096 steps STFT with a hop-length of 1024 and excluding the last bin for simplicity,
the input to the 5th layer of the spectral branch has only 8 frequency bins remaining, which are collapsed to a single one with a single convolution.
On the other hand, the input to the 5th layer of the temporal branch has an overall stride of $4^4=256$, which is aligned with the stride of 1024 of
the spectral branch with a single convolution. At this point, the two representations have the same shape and are summed before going through
a common layer that further reduces the number of time steps by 2. Symmetrically, the first layer in the decoder is shared before being fed both to
the temporal and spectral decoder, each with its own set of U-Net skip connections. The overall structure is represented in Fig.~\ref{fig:hybrid_demucs}.

The output of the spectrogram branch is inverted to a waveform, either directly with the ISTFT when CAC is used, or thanks to Wiener filtering~\cite{wiener}
for the amplitude representation, using Open-Unmix differentiable implementation~\cite{Stoter2019}. The final loss is applied directly in the time domain.
This allows for end-to-end hybrid domain training, with the model being free to combine both domains. To account for the fact that musical signals are not equivariant with respect to the frequency axis, we either inject a frequency embedding after the first spectral layer, following~\cite{zemb}, or we allow for different weights depending on the frequency band, as done by \cite{sony_densenet}.

Further improvements come from inserting residual branches in each of the encoder layers. The branches operate with a reduced number of dimensions
(scaled down by a factor of 4), using dilated convolutions and group normalization~\cite{groupnorm}, and for the 2 innermost layers, BiLSTM and local attention.
Local attention is based on regular attention~\cite{attention}, but replacing positional embedding with a controllable penalty limits its scope
to nearby time steps. All ReLUs in the network were replaced by GELUs~\cite{gelu}. Finally, we achieve better generalization and stability by penalizing
the largest singular values of each layer~\cite{spectral_norm}.
We achieved further gains (between 0.1 and 0.2dB) by fine-tuning the models on a specifically crafted dataset, and with longer training samples (30 seconds instead of 10). This dataset was built by combining stems from separate tracks, while respecting a number of conditions, in particular beat matching and pitch compatibility, allowing only for small pitch or tempo corrections.

The final models submitted to the competition are bags of 4 models. For leaderboard A, it is a combination of temporal only and hybrid Demucs models,
given that with only MUSDB18-HQ as a train set, we observed a regression on the bass source. For leaderboard B, all models are hybrid, as the extra training
data made the hybrid version better for all sources than its time-only version. We refer the reader to our Github repository {\color{purple}\href{https://github.com/facebookresearch/demucs}{facebookresearch/demucs}} for the exact hyper-parameters used.
More details and experimental results, including subjective evaluations, are provided in the hybrid Demucs paper~\cite{defossez2021hybrid}.

\subsection{KUIELab-MDX-Net (kuielab)} 
\label{ssec:kuielab}
\begin{figure*}
    \centering
    \includegraphics[width=0.8\textwidth]{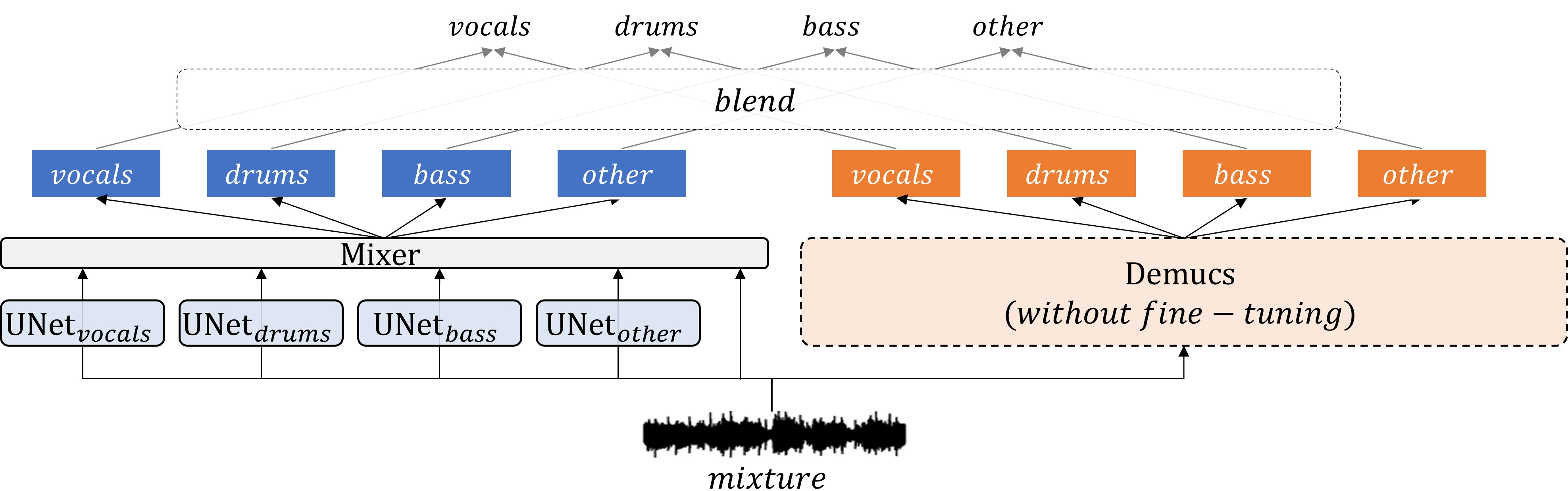}
    \caption{The Overall Architecture of KUIELab-MDX-Net.}
    \label{fig:mdx_net}
\end{figure*}

Similar to Hybrid Demucs, KUIELab-MDX-Net \cite{kim2021kuielab} uses a two-branched approach.
As shown in Fig.~\ref{fig:mdx_net}, it has a time-frequency branch (left-side) and a time-domain branch (right-side). 
Each branch estimates four stems (i.e., vocals, drums, bass, and other).
The \textit{blend} module \cite{Uhlich17} outputs the average of estimated stems from two branches for each source.
While branches of Hybrid Demucs were jointly trained end-to-end, each branch of KUIELab-MDX-Net was trained independently.

For the time-domain branch, it uses the original Demucs \cite{defossez2021music}, which was pre-trained on MUSDB18. 
It was not fine-tuned on MUSDB18-HQ, preserving the original parameters.

For the time-frequency-domain branch, it uses five sub-networks. Four sub-networks were independently trained to separate four stems, respectively. 
For each stem separation, an enhanced version of \textit{TFC-TDF-U-Net} \cite{choi_2020} was used.
We call the enhanced one \textit{TFC-TDF-U-Net v2} for the rest of the paper.
Another sub-network called \textit{Mixer} was trained to output enhanced sources by taking and refining the estimated stems.

TFC-TDF-U-Net is a variant of U-Net \cite{unet} architecture.
It improves source separation by employing TFC-TDF \cite{choi_2020} as building blocks instead of fully convolutional layers.
Architectural/training changes were made for TFC-TDF-U-Net v2 to the original as follows:

\begin{itemize}
    \item For skip connections between encoder and decoder, multiplication was used instead of concatenation.
    \item The other skip connections (e.g., dense skip connections in a dense block \cite{sony_densenet}) were removed.
    \item While the number of channels is not changed after down/upsampling in the original, channels are increased/decreased when downsampling/upsampling in v2.
    \item While the original was trained to minimize time-frequency domain loss, v2 was trained to minimize time-domain loss (i.e., $l_1$ loss between the estimated waveform and the ground-truth) 
\end{itemize}

    Since dense skip connections based on concatenation usually require a large amount of GPU memory, as discussed in \cite{chen2017}, TFC-TDF-U-Net v2 was designed to use simpler modules.
    \cite{kim2021kuielab} found that replacing concatenation with multiplication for each skip connection does not severely degrade the performance of the TFC-TDF-U-Net structure.
    Also, they observed that removing dense skip connections in each block does not significantly degrade the performance if we use TFC-TDF blocks.
    A single Time Distributed Fully connected (TDF) block, contained in a TFC-TDF, has an entire receptive field in the frequency dimension. Thus, U-Nets with TFC-TDFs can show promising results even with shallow or simple structures, as discussed in \cite{choi_2020}.
    To compensate for the lack of parameters by this design shift, it enlarges the number of channels for every downsampling, which is more general in the conventional U-Net \cite{unet}.
    It was trained to minimize time-domain loss for a direct end to end optimization.

Also, KUIELab-MDX-Net applies a frequency cut-off trick, introduced in \cite{Stoter2019}
to increase the window size of STFT (or \textit{fft size in short)
} as a source-specific preprocessing.
It cuts off high frequencies above the target source's expected frequency range from the mixture spectrogram. 
In this way, fft size could be increased while using the same input spectrogram size (which we needed to constrain for the separation time limit) for the model.
Since using a larger fft size usually leads to better SDR, this approach can improve quality effectively with a proper cut-off frequency. 
It is also why we did not use a multi-target model (i.e., a single model to separate all the sources once), where we could not apply source-specific frequency cutting.

Training one separation model for each source can have the benefit of source-specific preprocessing and model configurations.
However, these sub-networks lack the knowledge that they are separating using the same mixture because they cannot communicate with each other to share information.
An additional sub-network called \textit{Mixer} could further enhance the ``independently" estimated sources.
For example, estimated `vocals' often have drum snare noises left. The Mixer can learn to remove sounds from `vocals' that are also present in the estimated `drums' or vice versa.
Very shallow models (such as a single convolution layer) have been tried for the Mixer due to the time limit. 
One can try more complex models in the future, since even a single $1 \times 1$ convolution layer was enough to make some improvement on total SDR.
\textcolor{black}{
    Mixer used in KUIELab-MDX-Net is a point-wise convolution that is applied to the waveform domain.
    It takes a multi-channel waveform input containing four estimated stereo stem channels and the original stereo mixture channel. It outputs four different stereo stems.
    It can be viewed as a U-Net blending module with learnable parameters. An ablation study is provided in \cite{kim2021kuielab} for interested readers. 
}

Finally, KUIELab-MDX-Net takes the weighted average of estimated stems from two branches for each source. 
In other words, it blends \cite{Uhlich17} results from two branches.
For Leaderboard A, we trained KUIELab-MDX-Net which adopted TFC-TDF-U-Net v2, Mixer, and blending with the original Demucs, after training on MUSDB18-HQ \cite{MUSDB18HQ} training dataset with pitch/tempo shift \cite{defossez2021music} and random track mixing augmentation \cite{Uhlich17}. 
For Leaderboard B, we used KUIELab-MDX-Net without Mixer but with validation and test dataset of MUSDB18-HQ.
The source code for training KUIELab-MDX-Net is available at the Github repository {\color{purple}\href{https://github.com/kuielab/mdx-net}{kuielab/mdx-net}}.

\subsection{Danna-Sep (Kazane\_Ryo\_no\_Danna)}
\label{ssec:Kazane_Ryo_no_Danna}
\begin{figure*}
    \resizebox{\linewidth}{!}{\includegraphics{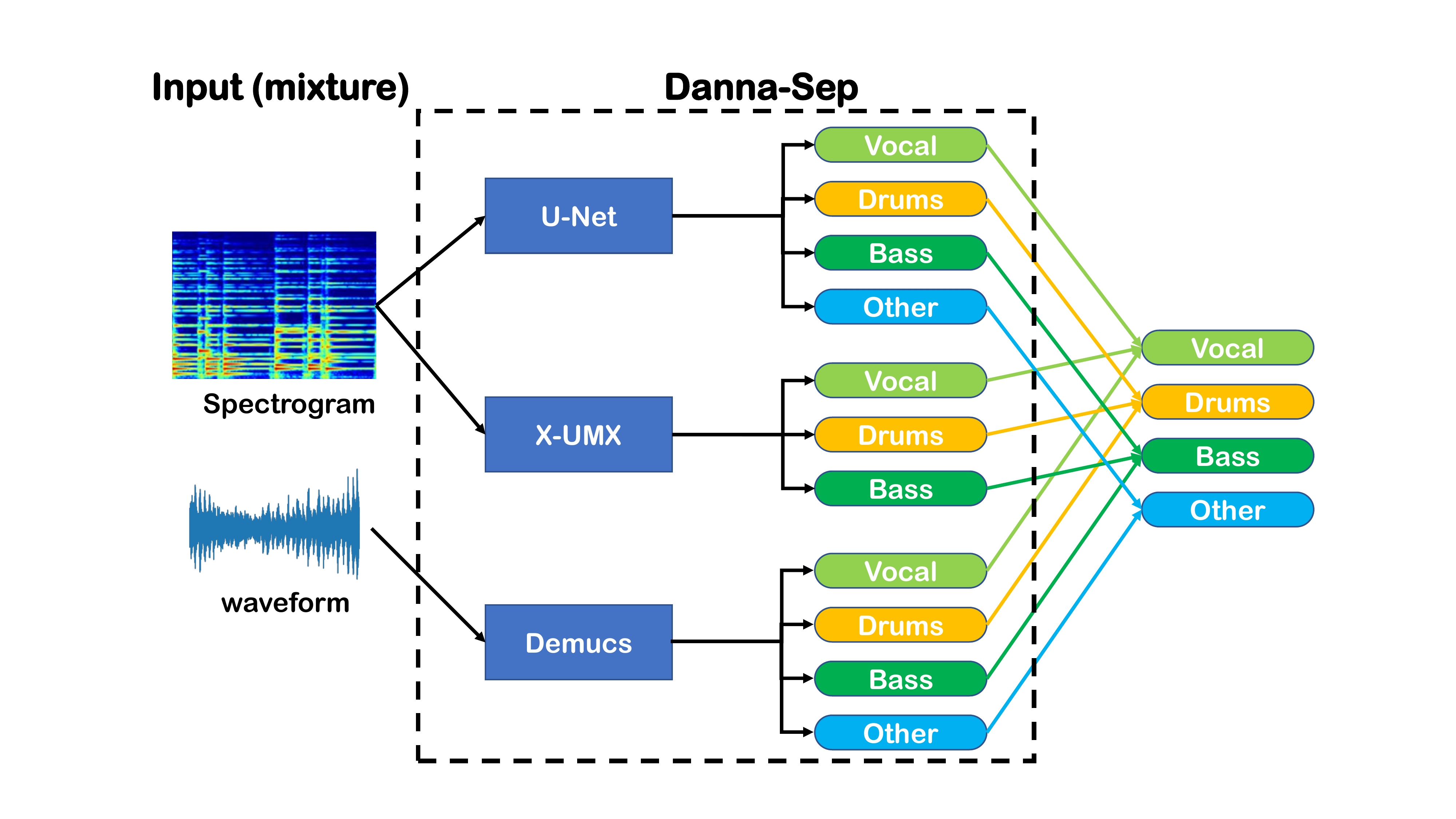}}
    \vspace{-1.0cm}
    \caption{The schematic diagram of Danna-Sep.}
    \label{fig:danna}
\end{figure*}

Danna-Sep is, not surprisingly, also a hybrid model blending the outputs from three source separation models across different feature domains. Two of them receive magnitude spectrogram as input, while the other one use waveforms, as shown in Fig.~\ref{fig:danna}. The design principle is to combine the complementary strengths of both waveform-based and spectrogram-based approaches. 

The first one of spectrogram-based models is a X-UMX trained with complex value frequency domain loss to better address the distance between ground truth and estimation spectrograms. In addition, we also incorporated differentiable Wiener filtering into training with our own PyTorch implementation \footnote{https://github.com/yoyololicon/norbert}, similar to how Hybrid Demucs did. We initialized this model with the official pre-trained weights \footnote{https://zenodo.org/record/4740378/files/pretrained\_xumx\_musdb18HQ.pth} before we start training. The second one is a U-Net with six layers consisting of D3 Blocks from D3Net~\cite{takahashi2021d3net} and two layers of 2D local attention~\cite{parmar2018image} at the bottleneck. We also experimented with using biaxial BiLSTM along the time and frequency axes as the bottleneck layers, but it took slightly longer to train yet offered a negligible improvement. We used the same loss function as our X-UMX during training but with Wiener filtering being disabled. 

The waveform-based model is a variant of 48 channels Demucs, where the decoder is replaced by four independent decoders responsible for four respective sources. We believe that the modification can help each decoder focusing on their target source without interfering with others. Each decoder has the same architecture as the original decoder, except for the size of the hidden channel which was reduced to 24. This makes the total number of parameters compared with the original Demucs. The training loss aggregates the L1-norm between estimated and ground-truth waveforms of the four sources. We didn't apply the shift trick~\cite{defossez2021music} to this variant because of limited computation resources set by the competition, but in our experiments, we found it still slightly outperformed the 48 channels Demucs. 

The aforementioned models were all trained separately on MUSDB18-HQ with pitch/tempo shift and random track mixes augmentation. 

Finally, we calculated the weighted average of individual outputs from all models. Experiments were conducted to search for optimal weighting. One iteration of Wiener filtering was used for our X-UMX and U-Net before averaging. To see the exact configurations we used, reader can refer to our repository  {\color{purple}\href{https://github.com/yoyololicon/music-demixing-challenge-ismir-2021-entry}{yoyololicon/music-demixing-challenge-ismir-2021-entry}} on Github. Experimental results of each individual model compare to the unmodified baselines are provided in our workshop paper \cite{yu2021dannasep}.

\section{Organizing the Challenge and Future Editions}
\label{Future}
\subsection{\textcolor{black}{General Considerations}}
Organizing a research challenge is a multi-faceted task: \textcolor{black}{the intention is to bring people from different research areas together, enabling them to tackle a task new to them in the easiest way possible; at the same time, }it is an opportunity for more experienced researchers to measure themselves once more against the state-of-the-art, providing them with new challenges and space to further improve their previous work.
All this, while making sure that the competition remains fair for all participants, both novice and experienced ones.

To satisfy the first point above, we decided to host the challenge on an open crowd-based machine learning competition website.
This allowed the competition to be visible by researchers outside of the source separation community.
The participants were encouraged to communicate, exchange ideas and even create teams, all through AIcrowd's forum.
To make the task accessible to everyone, we prepared introductory material and collected useful resources available on the Internet so that beginners could easily start developing their systems and actively take part in the challenge.

\textcolor{black}{We also increased the interest of experienced researchers because the new evaluation set represented an opportunity for them to evaluate their existing models on new data.}
This music was created for the specific purpose of being used as the test set in this challenge: this stimulated the interest of the existing MSS community because old models could be tested again and, possibly, improved.

The hidden test set also allowed us to organize a competition where fairness had a very high priority. Nobody could adapt their model to the test set as nobody except the organizers could access it.
The whole challenge was divided into two rounds, where different subsets of the test set were used for the evaluation: this prevented users from adapting their best model to the test data by running multiple submissions.
Only at the very end of the competition, the final scores on the complete test set were computed and the winners were selected: at that moment, no new model could be submitted anymore. Cash prizes are offered for the winners in return for open-sourcing their submission.

\subsection{Bleeding Sounds Among Microphones}
The hidden test set was created with the task of MSS in mind, which meant that we had to ensure as much diversity in the audio data as possible while maintaining a professional level of production and also deal with issues that are usually not present when producing music in a studio.

Depending on the genre of the music being produced, bleeding among different microphones (i.e., among the recorded tracks of different instruments) is more or less tolerated. For example, when recording an orchestra, to maintain a natural performance, the different sections play in the same room: even if each section may be captured with a different set of microphones, each microphone will still capture the sound of all the other sections.
This phenomenon is not desirable for the task of MSS: if a model learns to separate each instrument very well but the test data contains bleeding, the model will be wrongly penalized during the evaluation.
For this reason, when producing the songs contained in the dataset, we had to ensure that specific recording conditions were respected. All our efforts were successful, except for one song (SS\_008): even taking appropriate measures during the recording process (e.g., placing absorbing walls between instruments), the tracks contained some bleeding between drums and bass. For this reason, we removed this track from the evaluation and used it as a demo song instead.

Bleeding can be an issue also for training data: we cannot expect models to have better performance than the data they are trained upon unless some specific measures are taken.
We did not explicitly address this aspect in the MDX Challenge; nevertheless, designing systems that are robust to bleeding in the training data is a desirable feature, not just for MSS.
We could envision that future editions of the MDX Challenge will have a track reserved for systems trained exclusively on data that suffer from bleeding issues, to focus the research community on this aspect as well.

\subsection{Silent Sources}
Recording conditions are not the only factor that can make a source separation evaluation pipeline fail.
Not all instruments are necessarily present in a track: depending on the choices of the composer, arranger, or producer, some instruments may be missing.
This is an important aspect, as some evaluation metrics, like the one we chose, may not be robust to the case of a silent target.
We decided to exclude from MDXDB21 one song (SS\_015) as it has a silent target for the bass track.
Please note that this issue was not present in previous competitions like SiSEC, as the test set of MUSDB18 does not feature songs with silent sources.
Nevertheless, we think it would be an important improvement if the challenge evaluation could handle songs where one of the instruments is missing (e.g., instrumental songs without vocals, acapella interpretations, etc.).

Such an issue arises from the definition of clear identities for the targets in the source separation task: the evaluation pipeline suffers from such strict definition and causes some songs to be unnecessarily detrimental to the overall scores.
This is a motivation to move towards source separation tasks that do not feature clear identities for the targets, such as universal sound source separation: in that case, the models need to be able to separate sounds, independently on their identity.
An appropriately designed pipeline for such a task would not suffer from the issue above.
For this reason, for future editions of the challenge, we may consider including tasks similar to universal sound source separation.

\subsection{Future Editions}
We believe that hosting the MDX Challenge strengthened and expanded the source separation community, by providing a new playground to do research and attracting researchers from other communities and areas, allowing them to share knowledge and skills: all this focused on solving one of the most interesting research problems in the area of sound processing.
Source separation can still bring benefits to many application and research areas: this motivates the need for future editions of this competition.

In our view, this year's edition of the MDX Challenge allows us to start a hopefully long tradition of source separation competitions. The focus of this year was MSS on four instruments: given the role this task played in past competitions, this was a convenient starting point, that provided us with enough feedback and experience on how to make the competition grow and improve.

We encountered difficulties when compromising between how the source separation task expects data to be created and the professional techniques for music production: for instance, to keep the competition fair, we had to make sure that no crosstalk between target recordings was present.
The same argument about crosstalk also highlighted the need for source separation systems that can be trained on data that suffer from this issue: this potentially opens access to training material not available before and can be another source of improvement for existing models.
We realized how brittle the design of a simple evaluation system is when dealing with the vast amount of choices that artists can make when producing music: even the simple absence of an instrument in a track can have dramatic consequences in the competition results.

This knowledge will eventually influence the decisions we will take when designing the next editions of the MDX Challenge. In particular, we will:
\begin{itemize}
    \item design an evaluation system around a metric that is robust to bleeding sounds between targets in the test ground truth data
    \item direct the attention of researchers to the robustness of models concerning bleeding among targets in the training data, possibly reserving a separate track to systems trained exclusively on such data
    \item partially moves towards source separation tasks where there is no predefined identity for the targets, such as universal sound source separation.
\end{itemize}

Furthermore, motivated by the pervasiveness of audio source separation, we will consider reserving special tracks to other types of audio signals, such as speech and ambient noise.
The majority of techniques developed nowadays provide useful insights independently on whether they are applied to music, speech, or other kinds of sound.
In the interest of scientific advancement, we will try to make the challenge as diverse as possible, to have the highest number of researchers cooperate, interact and ultimately compete for the winning system.

\section{Conclusions}
With the \emph{MDX Challenge 2021}, we continued the successful series of SiSEC MUS challenges. By using a crowd-sourced platform to host the competition, we tried to make it easy for ML practitioners from other disciplines to enter this field. Furthermore, we introduced a newly created dataset, called MDXDB21, which served as the hidden evaluation set for this challenge. Using it allows to fairly compare all recently published models as it shows their generalization capabilities towards unseen songs. 

We hope that this MDX Challenge will be the first one in a long series of competitions.

\bibliography{MDX_Challenge_2021_Frontier}

\appendix
\newpage
\section{Additional information for MDXDB21}
\label{appendix:AdditionalInformation}

Table~\ref{tab:statistics_mdxdb21} provides more information about the songs in MDXDB21, which allows a better interpretation of the SDR results that the participants obtained.
In particular, Table~\ref{tab:statistics_mdxdb21} contains for each song as well as its stems the following statistics:
\begin{itemize}
    \item Maximum absolute peak value (left/right channel),
    \item Loudness according to BS.1770 (left/right channel),
    \item Correlation coefficient between left/right channel (time-domain).
\end{itemize}

\begin{table*}
    \centering
    \resizebox{\linewidth}{!}{%
    \begin{tabular}{|c|c|c|c|c|c|c|c|c|c|c|c|c|c|c|c|}
    \hline
    \multirow{2}{*}{\textbf{Song ID}} & \multicolumn{3}{c}{\textbf{Mixture}} & \multicolumn{3}{|c|}{\textbf{Bass}} & \multicolumn{3}{|c|}{\textbf{Drums}}& \multicolumn{3}{|c|}{\textbf{Other}}& \multicolumn{3}{|c|}{\textbf{Vocals}}\\
    & Max-Abs  & Loudness & Corr.
    & Max-Abs  & Loudness & Corr.
    & Max-Abs  & Loudness & Corr.
    & Max-Abs  & Loudness & Corr.
    & Max-Abs  & Loudness & Corr. \\
    \hline
    \hline
SS\_001  & 0.77, 0.80 & -20.3, -20.2 & 0.74 & 0.17, 0.17 & -30.7, -30.7 & 1.00 & 0.53, 0.53 & -27.5, -27.6 & 0.99 & 0.43, 0.43 & -24.4, -24.2 & 0.41 & 0.57, 0.60 & -21.2, -21.2 & 0.93 \\
\hline
SS\_002  & 0.86, 0.87 & -19.3, -19.3 & 0.89 & 0.31, 0.31 & -25.2, -25.2 & 1.00 & 0.45, 0.45 & -25.1, -24.5 & 0.98 & 0.45, 0.42 & -25.1, -25.7 & 0.59 & 0.44, 0.43 & -22.5, -22.6 & 0.90 \\
\hline
SS\_003  & 0.85, 0.69 & -20.3, -21.4 & 0.64 & 0.15, 0.15 & -28.5, -28.5 & 1.00 & 0.23, 0.22 & -35.7, -35.6 & 0.82 & 0.44, 0.42 & -26.8, -27.5 & 0.12 & 0.68, 0.58 & -21.1, -22.4 & 0.82 \\
\hline
SS\_004  & 0.88, 0.94 & -18.2, -18.6 & 0.77 & 0.35, 0.35 & -29.6, -29.6 & 1.00 & 0.70, 0.67 & -25.0, -25.6 & 0.81 & 0.52, 0.65 & -23.9, -24.5 & 0.47 & 0.67, 0.63 & -21.1, -21.2 & 0.85 \\
\hline
SS\_005  & 0.84, 0.91 & -19.8, -20.1 & 0.74 & 0.23, 0.23 & -28.5, -28.5 & 1.00 & 0.52, 0.56 & -29.7, -29.3 & 0.94 & 0.40, 0.46 & -26.0, -26.3 & 0.10 & 0.71, 0.63 & -21.0, -21.3 & 0.93 \\
\hline
SS\_006  & 0.99, 0.95 & -18.5, -18.7 & 0.78 & 0.30, 0.28 & -25.2, -25.5 & 0.98 & 0.45, 0.42 & -28.1, -28.0 & 0.93 & 0.52, 0.57 & -23.4, -23.7 & 0.32 & 0.49, 0.49 & -21.1, -21.0 & 0.94 \\
\hline
SS\_007  & 0.95, 0.87 & -18.2, -18.4 & 0.88 & 0.32, 0.32 & -23.0, -23.0 & 1.00 & 0.61, 0.54 & -24.7, -24.7 & 0.91 & 0.55, 0.62 & -22.1, -22.8 & 0.53 & 0.42, 0.44 & -23.0, -22.8 & 0.96 \\
\hline
SS\_008  & 0.86, 0.84 & -18.3, -18.2 & 0.78 & 0.31, 0.27 & -25.9, -26.3 & 0.96 & 0.46, 0.43 & -25.4, -26.2 & 0.82 & 0.43, 0.50 & -24.6, -23.5 & 0.49 & 0.43, 0.43 & -21.1, -21.3 & 0.87 \\
\hline
SS\_009  & 0.93, 0.96 & -18.7, -18.6 & 0.62 & 0.19, 0.19 & -26.4, -26.4 & 1.00 & 0.75, 0.70 & -27.2, -27.6 & 0.84 & 0.50, 0.50 & -22.8, -22.9 & 0.35 & 0.54, 0.49 & -21.9, -21.6 & 0.58 \\
\hline
SS\_010  & 0.87, 0.81 & -18.9, -18.8 & 0.64 & 0.45, 0.43 & -25.2, -25.7 & 0.55 & 0.50, 0.51 & -22.7, -22.5 & 0.95 & 0.45, 0.44 & -24.8, -24.5 & 0.05 & 0.29, 0.29 & -25.0, -25.0 & 0.88 \\
\hline
SS\_011  & 0.83, 0.97 & -17.3, -17.0 & 0.72 & 0.26, 0.26 & -24.3, -24.5 & 1.00 & 0.45, 0.50 & -22.7, -21.4 & 0.77 & 0.64, 0.53 & -21.5, -21.9 & 0.35 & 0.39, 0.40 & -23.1, -23.1 & 0.88 \\
\hline
SS\_012  & 0.91, 0.94 & -18.5, -18.5 & 0.75 & 0.29, 0.27 & -28.8, -28.9 & 0.98 & 0.49, 0.48 & -30.0, -30.0 & 0.80 & 0.54, 0.57 & -24.3, -24.3 & 0.28 & 0.63, 0.63 & -19.1, -19.1 & 0.96 \\
\hline
SS\_013  & 0.68, 0.72 & -20.0, -20.1 & 0.66 & 0.20, 0.20 & -29.6, -29.6 & 1.00 & 0.46, 0.40 & -27.5, -27.6 & 0.95 & 0.52, 0.49 & -23.1, -23.4 & 0.36 & 0.41, 0.40 & -24.6, -24.5 & 0.78 \\
\hline
SS\_014  & 0.90, 0.89 & -19.8, -19.8 & 0.77 & 0.27, 0.25 & -24.9, -25.0 & 0.95 & 0.72, 0.83 & -25.4, -24.6 & 0.91 & 0.37, 0.36 & -25.5, -26.7 & 0.02 & 0.50, 0.54 & -23.1, -23.1 & 0.96 \\
\hline
SS\_015  & 0.87, 0.99 & -20.4, -19.9 & 0.68 & 0.00, 0.00 & -inf, -inf & 0.00 & 0.63, 0.79 & -24.9, -23.6 & 0.62 & 0.51, 0.51 & -24.9, -24.6 & 0.62 & 0.40, 0.42 & -23.9, -23.7 & 0.82 \\
\hline
SS\_016  & 0.65, 0.66 & -20.6, -20.2 & 0.64 & 0.26, 0.26 & -26.2, -26.1 & 0.99 & 0.40, 0.38 & -27.8, -28.2 & 0.95 & 0.29, 0.31 & -25.6, -25.1 & 0.08 & 0.46, 0.42 & -26.0, -25.2 & 0.43 \\
\hline
SS\_017  & 0.72, 0.72 & -22.7, -22.6 & 0.98 & 0.15, 0.17 & -30.8, -29.9 & 1.00 & 0.57, 0.60 & -28.5, -28.2 & 0.99 & 0.38, 0.39 & -28.1, -28.1 & 0.97 & 0.48, 0.41 & -27.0, -27.3 & 0.93 \\
\hline
SS\_018  & 0.90, 0.92 & -18.9, -18.9 & 0.84 & 0.44, 0.44 & -23.2, -23.2 & 1.00 & 0.48, 0.48 & -23.7, -23.6 & 0.95 & 0.43, 0.47 & -24.5, -24.6 & 0.37 & 0.35, 0.34 & -26.3, -26.4 & 0.91 \\
\hline
SS\_019  & 0.92, 0.99 & -18.1, -18.1 & 0.81 & 0.16, 0.16 & -26.9, -27.0 & 1.00 & 0.49, 0.48 & -21.4, -21.9 & 0.97 & 0.37, 0.47 & -26.1, -25.3 & 0.19 & 0.46, 0.47 & -22.3, -22.1 & 0.77 \\
\hline
SS\_020  & 0.91, 0.98 & -20.7, -21.1 & 0.74 & 0.26, 0.26 & -25.5, -25.7 & 1.00 & 0.72, 0.76 & -26.3, -26.2 & 0.95 & 0.38, 0.39 & -23.9, -24.8 & 0.24 & 0.22, 0.21 & -27.7, -27.6 & 0.95 \\
\hline
SS\_021  & 0.93, 0.97 & -18.8, -18.7 & 0.80 & 0.39, 0.39 & -23.9, -23.8 & 1.00 & 0.59, 0.55 & -24.2, -24.0 & 0.91 & 0.52, 0.60 & -25.1, -25.2 & -0.07 & 0.40, 0.38 & -22.6, -22.6 & 0.87 \\
\hline
SS\_022  & 0.92, 0.94 & -17.2, -17.3 & 0.79 & 0.28, 0.30 & -24.6, -24.6 & 1.00 & 0.48, 0.45 & -22.6, -22.7 & 0.94 & 0.41, 0.43 & -22.2, -22.8 & 0.28 & 0.75, 0.74 & -21.4, -21.4 & 0.92 \\
\hline
SS\_023  & 0.83, 0.85 & -18.4, -18.2 & 0.77 & 0.28, 0.27 & -26.6, -26.4 & 0.97 & 0.61, 0.58 & -25.1, -25.3 & 0.83 & 0.34, 0.40 & -24.4, -24.5 & 0.46 & 0.74, 0.74 & -21.1, -21.0 & 0.81 \\
\hline
SS\_024  & 0.91, 0.95 & -17.2, -17.2 & 0.69 & 0.42, 0.45 & -22.7, -22.5 & 0.70 & 0.81, 0.82 & -21.6, -21.9 & 0.92 & 0.45, 0.46 & -24.5, -24.2 & 0.02 & 0.56, 0.57 & -21.8, -21.8 & 0.84 \\
\hline
SS\_025  & 0.91, 0.91 & -21.9, -21.9 & 0.82 & 0.25, 0.25 & -28.0, -28.0 & 1.00 & 0.41, 0.43 & -28.6, -28.3 & 0.86 & 0.57, 0.56 & -26.3, -26.2 & 0.68 & 0.37, 0.37 & -24.9, -25.2 & 0.85 \\
\hline
SS\_026  & 0.98, 0.79 & -20.8, -21.2 & 0.70 & 0.49, 0.48 & -26.2, -26.2 & 0.84 & 0.63, 0.64 & -26.1, -26.0 & 0.88 & 0.80, 0.50 & -24.0, -25.0 & 0.38 & 0.13, 0.14 & -28.6, -28.6 & 0.74 \\
\hline
SS\_027  & 0.64, 0.70 & -21.0, -21.0 & 0.57 & 0.19, 0.19 & -29.9, -29.9 & 1.00 & 0.43, 0.36 & -25.6, -25.6 & 0.94 & 0.50, 0.51 & -23.6, -23.7 & 0.27 & 0.24, 0.25 & -26.1, -26.1 & 0.63 \\
\hline
SS\_028  & 0.59, 0.61 & -23.2, -23.0 & 0.85 & 0.14, 0.14 & -32.6, -32.6 & 1.00 & 0.43, 0.47 & -27.0, -26.5 & 0.96 & 0.18, 0.24 & -32.3, -31.8 & 0.39 & 0.30, 0.25 & -27.0, -27.1 & 0.69 \\
\hline
SS\_029  & 0.66, 0.63 & -21.0, -20.9 & 0.66 & 0.25, 0.26 & -28.7, -28.9 & 0.99 & 0.36, 0.40 & -25.7, -25.6 & 0.96 & 0.37, 0.38 & -25.8, -25.6 & 0.09 & 0.33, 0.30 & -26.4, -26.6 & 0.73 \\
\hline
SS\_030  & 0.73, 0.76 & -18.3, -18.7 & 0.71 & 0.23, 0.23 & -26.0, -26.1 & 1.00 & 0.38, 0.40 & -24.3, -24.7 & 0.85 & 0.44, 0.40 & -22.7, -23.3 & 0.33 & 0.40, 0.37 & -24.0, -24.3 & 0.71 \\
\hline
    \end{tabular}}
    \caption{Statistics about songs and their stems in MDXDB21. ``Max-Abs'' and ``Loudness'' values are given for the left and right channel, respectively.}
    \label{tab:statistics_mdxdb21}
\end{table*}

\end{document}